# Spatiotemporal patterns and predictability of cyberattacks

Yu-Zhong Chen[1], Zi-Gang Huang[1,2*], Shouhuai Xu[3], Ying-Cheng Lai[1,4*]

1 School of Electrical, Computer and Energy Engineering, Arizona State University, Tempe, Arizona 85287, USA

2 Institute of Computational Physics and Complex Systems, Lanzhou University, Lanzhou Gansu 730000, China

3 Department of Computer Science, University of Texas at San Antonio, San Antonio, TX 78249, USA

4 Department of Physics, Arizona State University, Tempe, Arizona 85287, USA

∗ Email: huangzg@lzu.edu.cn (ZGH); Ying-Cheng.Lai@asu.edu (YCL)

## Abstract

A relatively unexplored issue in cybersecurity science and engineering is whether there exist intrinsic patterns of cyberattacks. Conventional wisdom favors absence of such patterns due to the overwhelming complexity of the modern cyberspace. Surprisingly, through a detailed analysis of an extensive data set that records the time-dependent frequencies of attacks over a relatively wide range of consecutive IP addresses, we successfully uncover intrinsic spatiotemporal patterns underlying cyberattacks, where the term "spatio" refers to the IP address space. In particular, we focus on analyzing *macroscopic* properties of the attack traffic flows and identify two main patterns with distinct spatiotemporal characteristics: deterministic and stochastic. Strikingly, there are very few sets of major attackers committing almost all the attacks, since their attack "fingerprints" and target selection scheme can be unequivocally identified according to the very limited number of unique spatiotemporal characteristics, each of which only exists on a consecutive IP region and differs significantly from the others. We utilize a number of quantitative measures, including the flux-fluctuation law, the Markov state transition probability matrix, and predictability measures, to characterize the attack patterns in a comprehensive manner. A general finding is that the attack patterns possess high degrees of predictability, potentially paving the way to anticipating and, consequently, mitigating or even preventing large-scale cyberattacks using macroscopic approaches.

# Introduction

Highly networked communication and information infrastructures built via various state-of-the-art technologies play crucial roles in modern economic, social, military, and political activities. However, such sophisticated infrastructures are facing more and more severe security challenges on the global scale [1–4]. Earlier theoretical works focused on understanding the complex topologies of the Internet [5] and on the likelihood of large scale failures caused by node removal in complex networks [6–10]. Recent years have witnessed tremendous efforts devoted to mitigating and coping with increasing cybersecurity threats. For example, attack graphs were invented to analyze the overall network vulnerability and to generate a global view of network security against attacks [11–14]. By deploying network sensors at particular points in the Internet, monitoring systems were built to detect cyberthreats and statistically analyze the time, sources, and the types of attacks [15], and various visualization methods were developed to better understand the result of the detection and analysis [16–19]. Quite recently, a genetic epidemiology approach to cybersecurity was proposed to understand the factors that determine the likelihood that individual computers are compromised [20], and the general concept of cybersecurity dynamics was introduced [21].

Attack traffic analysis were mainly done in the field of Intrusion Detection System (IDS), the cyberspace's equivalent to the burglar alarm. IDS has become one of the fundamental technologies for network security [22]. There are three approaches to building an IDS: (1) signature or misuse detection, (2) anomaly detection, and (3) hybrid or compound detection. In particular, signature detection technique is based on a predefined set of known attack signatures obtained from security experts. The system observes the activities of subjects and alarms if their behaviors match the malicious ones in the attack signature set. Both host-based [23, 24] and network-based [25, 26] detection systems were developed. Anomaly detection technique is based on machine learning methodologies, such as system call based sequence analysis [27–29], Bayesian networks [30–32], principal component analysis [33–35], and Markov models [36, 37]. The IDS monitoring capability can be improved by taking a hybrid approach that combines both signature and anomaly detection strategies [38,39]. All these methods are often based on data packet payload inspection and thus are difficult to perform for high speed networks. Another limitation of these approaches is the assumption that either the attacks are well defined (i.e., signatures) or the normal behaviors are well defined (so are the abnormal behaviors). Recently, there has been a growing interest in flow-based intrusion detection technologies, by which communication patterns within the network are analyzed, instead of the contents of individual packets [40, 41]. Interestingly, a quite recent study analyzing the data obtained from the host IDSs reveals strong associations between the network services running on the host and the specific types of threats to which it is susceptible [20]. Making use of the plan recognition method in artificial intelligence, one can predict the attack plan from the IDS alert information [42]. Utilizing virtual or physical networks to test these IDS techniques can be costly and time consuming, hence, as an alternative, simulation modeling approaches were developed to represent computer networks and IDS to efficiently simulate cyberattack scenarios [43–45]. As botnets have become a major threat in cyberspace, cyberattack traffic patterns have also been used to understand botnet's Command-and-Control strategies [46–48].

In this paper, we uncover the existence of intrinsic spatiotemporal patterns underlying cyberattacks and address the important question of whether certain such attacks may be predicted or anticipated in advance. The overwhelming complexity of the modern cyberspace would suggest complete randomness in the distribution of cyberattacks and, as a result, the intuitive expectation is that attackers' behaviors are random and attacks are unpredictable. However, our discovery of the spatiotemporal patterns and quantitative characterization of the predictability of these patterns suggest the otherwise. In particular, distinct from previous works on cyberattack analysis, our efforts concentrate on analyzing the *macroscopic* properties of the attack traffic flows using a data set of cyberattacks available to us. Especially, the data set recorded attacks on 491 consecutive victim IP addresses (sensors) in 18 days. The IP addresses can thus be regarded, approximately, as a variable in space. An attack is regarded as an event occurring in both space and time, and we speak of events in spatiotemporal dimensions. This is much more comprehensive than the analysis of the individual time series obtained from sampled IP addresses or the time series obtained by treating the IP addresses as a whole [49–57] Our results reveal, for the first time, that robust macroscopic patterns exist in the seemingly random cyberspace: majority of the attacks are governed by a few very limited number of patterns, indicating that cyberattacks are mainly committed by a few types of major attackers, each with unique spatiotemporal characteristics. More specifically, the patterns can be divided into two types: deterministic and stochastic. The emergence of deterministic patterns implies predictability, which can potentially be exploited to anticipate certain types of attacks to achieve greater cybersecurity. We characterize the predictability of attack frequency time series based on information entropy [58]. Our results suggest a surprisingly high degree of predictability, especially for the IPs under deterministic attack. Effective algorithms can then be developed to predict the future attack frequencies. We also develop methods to evaluate the inference probability between the attack frequency time series based on series similarity, which may allow us to plant much fewer attack probes into the Internet while still achieving effective monitoring. The stochastic patterns can be quantified using the flux-fluctuation law in statistical and nonlinear physics [59–68]. Our findings outline a global picture of how cyberattacks are initiated and distributed into the Internet. This will be of potential value to the development of defense strategies against cyberattacks on a global scale.

## Results

*Existence of IP address blocks sharing similar attack patterns and common backgrounds*. We define attack frequency $w(t)$ as the number of attacks received by a victim IP per time unit $\Delta t$. Figure 1 shows the time series $w(t)$ from all the victim IP addresses for $\Delta t = 1000$ seconds. Surprisingly, instead of overwhelming randomness, we observe substantial regularity: the IP-space can be unequivocally divided into distinct colored blocks, where the amplitudes of the time series within each block are approximately of the same order of magnitude, but the amplitudes from different blocks vary considerably. Note that, within each block, the time series $w(t)$ are approximately synchronized, which correspond to a particular attack pattern. For different attack patterns, there exists a common background in the different IP sectors. For example, about half of the IP addresses (from 1 to 246) belong to the background of aqua block $[w(t) \sim 10^{1.5}]$, in which yellow, orange, and red patterns all exist, while all other IP addresses possess a dark-blue background $[w(t) < 10^{0.5}]$ with a number of vertical light-blue lines going through them.

To further investigate the properties of each pattern, we concentrate on a small time period, as shown in Fig. 2. Firstly, under much higher time resolution ($\Delta t = 10$sec), the large number of vertical light-blue lines in Fig. 1 (a) become curves evenly distributed along the $t$ axis and occupying about half of the IP-space (from 1 to 246). These curves are approximately parallel lines within the same IP region but with different slopes in different IP regions (the sloped lines become visually vertical under low time resolution or large observational time period). It is these curves that form the aqua background throughout the upper half of the IP space in Fig. 1. Secondly, all the light-blue curves exist on a dark-blue background with three vertical light-blue lines appearing throughout the entire IP-space, indicating a common background covering every single IP address on which other attack patterns are superimposed. We call the vertical lines "walls," which represent the level of simultaneous attacks on each victim IP and are visually similar to walls in the three-dimensional representation, as shown in Figs. 1(b) and 2(b).

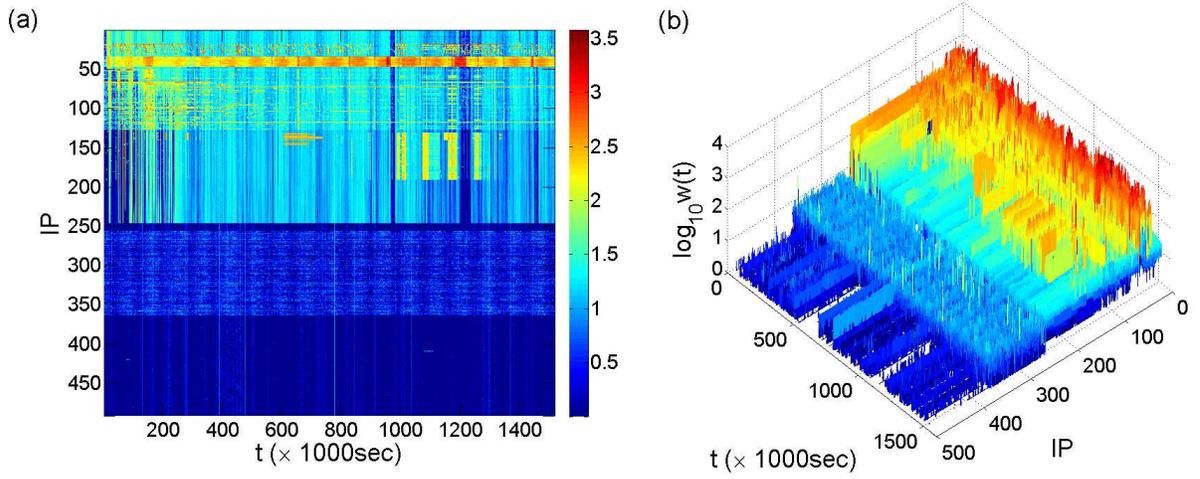

Figure 1: **Time series of attack frequency** $w(t)$ **for all IP addresses.** (a) Spatiotemporal representation of attack frequencies $w(t)$ of all IP addresses on a logarithmic scale, where the $x$-axis and $y$-axis are, respectively, time $t$ and IP address index from 1 to 491 (top to bottom). The IP region 1-246 is denoted as the aqua background, where each of the four IP regions (19-31, 35-47, 50-130, and 131-191) exhibits a particular attack pattern that is overlayed on the background. The IP region 247-363 possesses an attack pattern that is overlayed on the dark-blue background lying under the entire IP-space. (b) Three-dimensional presentation of the attack frequency $w(t)$ on a logarithmic scale. The "walls" of unity height are not visible on the logarithmic scale, but higher "walls" are visible in the IP region 364-491 [corresponding to the vertical light-blue lines on the bottom dark-blue background in (a)], which actually occur in the entire IP-space but are mixed with other patterns. The time resolution is $\Delta t = 1000$ seconds.

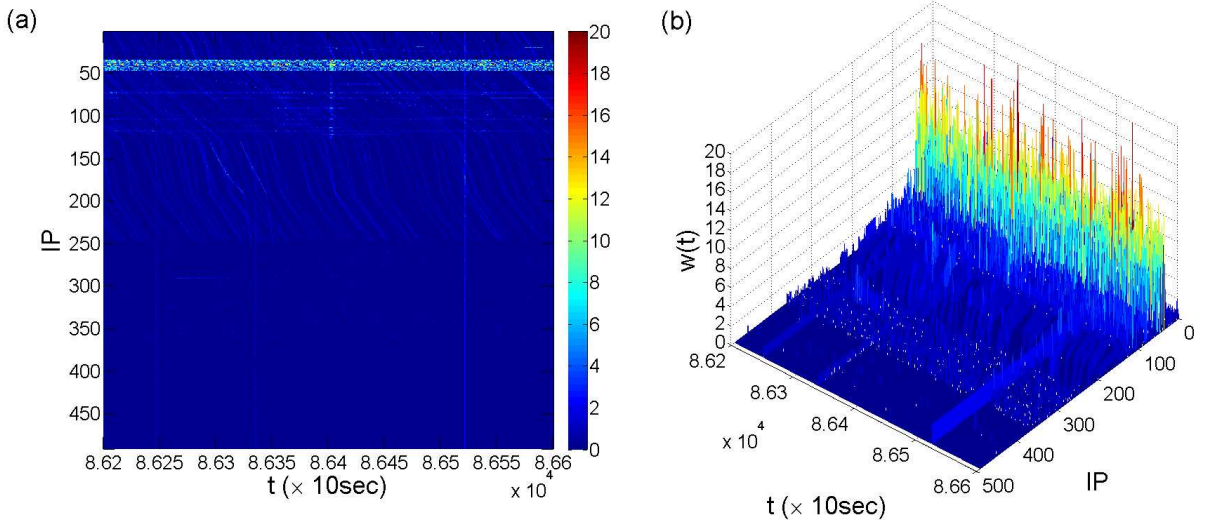

Figure 2: **Time series of attack frequency with higher time resolution.** Time range: $t = 86200$ to 86600. Time resolution: $\Delta t = 10$ seconds. (a,b) Two- and three-dimensional representations of the attack frequency $w(t)$ for the entire IP-space on a linear scale, respectively.

These observations have the following implications. Firstly, IP addresses in the same colored block are attacked through quite similar patterns, and the block as a whole shows unique spatial-temporal features, which can be effectively associated with the "fingerprint" of a particular set of attackers. For an IP block corresponding to a particular attack pattern, since the relatively strong spatiotemporal regularity of the pattern can only be achieved via attacks highly organized in time and space, it is reasonable to speculate that the corresponding attacks are solely committed by a set of intimately correlated attackers, or even one major attacker, rather than by a set of multiple independent attackers. A single major attacker is more likely to play the role, due to the requirement of close cooperation between the attackers that may be unrealistic. Thus, a particular type of attack patterns, or fingerprints, corresponds to a particular major attacker. Globally, according to the limited types of attack patterns, a small number of major attackers are responsible for almost all the attacks. Secondly, the consecutiveness of IP addresses attacked in similar patterns reveals the way how the corresponding attackers select their targets, i.e., targeting each IP address within a consecutive IP sector rather than distantly separated IP addresses. Thirdly, the coexistence of multiple patterns overlaying on some identical background indicates that some IP addresses may be under the cross fire from multiple major attackers.

<u>Deterministic attacks.</u> Counterintuitively, the parallel light-blue lines in Fig. 2(a) constituting the aqua background in the upper half IP-space show a substantial component of deterministic attack behaviors. We speculate that each light-blue line is generated by one of the attacker's devices, such as a zombie computer in a botnet that launches attacks on a certain IP region with a constant sweeping speed and a certain order with similar time intervals. Such an "organized" attack pattern makes the light-blue lines nearly evenly distributed along the time axis, which can be regarded as *deterministic attacks*. Because of the deterministic rules that the attacks follow, it is possible to predict when and where the next attack is going to take place by identifying the ordered time interval and extrapolating the sweeping speed. We observe that the attacker associated with a light-blue line typically attacks each IP address once approximately within every 3 to 8 seconds. Specifically, the three different sweeping rates in the three sub-regions of IP addresses are about 8 seconds per IP from 51 to 130, 3 seconds per IP from 131 to 191, and 6 seconds per IP from 192 to 246. Similar deterministic attack patterns can also be observed in a relatively narrow IP region close to the top of the IP region (IP 19-31), but in a much larger time scale, indicated as the thicker light-green lines right below the top of Fig. 3(a). Fig. 3(b) is the enlarged graph with the contrast slopes of the two deterministic attack patterns shown more clearly. The deterministic attacks in the IP range 19-31 typically take about 200 to 1100 seconds per IP before switching to the next, with the attack rate of about once per second.

The "wall" attacks in the dark-blue background show another type of deterministic attack patterns [Figs. 1 and 2], which are instantaneous attacks to each IP that can be observed with time resolution of $\Delta t = 1$ second. However, when a higher time resolution is used, e.g., $\Delta t = 10^{-3}$ second, the "wall" attacks are found to occur one after another in the order of the IP index. Fig. 4 shows two typical types of "wall" attacks. For the first type, attacks are performed exactly once on each IP in the order of IP index but with time delays of the order of $10^{-2}$ second. This is the most frequently observed walls with unity height in the data set, while close occurrence of several "wall" attacks induces a higher wall when larger values of $\Delta t$ are used. The second kind of "wall" attacks shown in Figs. 4(b) and 4(c) consists of 5 consecutive attacks on one IP before skipping to the next, strictly in terms of the IP order.

The three different deterministic attack patterns observed in Fig. 3, i.e., the thick green lines in IP 19-31, the background in IP 1-246, the "wall" attacks over the whole IP space, all sweep the IP space strictly in order but with time scales differing by orders of magnitude. The multiscale behaviors associated with the deterministic attack patterns are crucial for understanding and predicting attacks in the cyberspace.

*Flux-fluctuation relation and stochastic attack.* Recently there has been a great deal of attention to the flux-fluctuation relations in complex networked systems with stochastic transportation dynamics [59–68], where the flux $w(t)$ of a node is the amount of goods or the number of data packets that it transmits per time unit at time $t$, and the standard deviation $\sigma$ of $w(t)$ corresponds to the fluctuation. The following flux-fluctuation relation holds when the traffic system evolves under a given external drive [66]:

$$\sigma = \sqrt{\langle w \rangle + \left( \frac{\sigma_{ext}^2}{\langle W \rangle^2} - \frac{1}{\langle W \rangle} \right) \cdot \langle w \rangle^2}, \tag{1}$$

where $\langle w \rangle$ is the time average of flux $w(t)$, $\langle W \rangle$ is the time average of the external drive $W(t)$, and $\sigma_{ext}$ denotes the standard deviation of $W(t)$. If the external drive $W$ is approximately a constant, i.e., $\sigma_{ext} \approx 0$, or if $W$ follows the Poisson distribution ($\sigma_{ext} = \langle W \rangle$), then $\sigma \sim \sqrt{\langle w \rangle}$. However, if the external drive has large fluctuations, the relation becomes $\sigma \sim \langle w \rangle$.

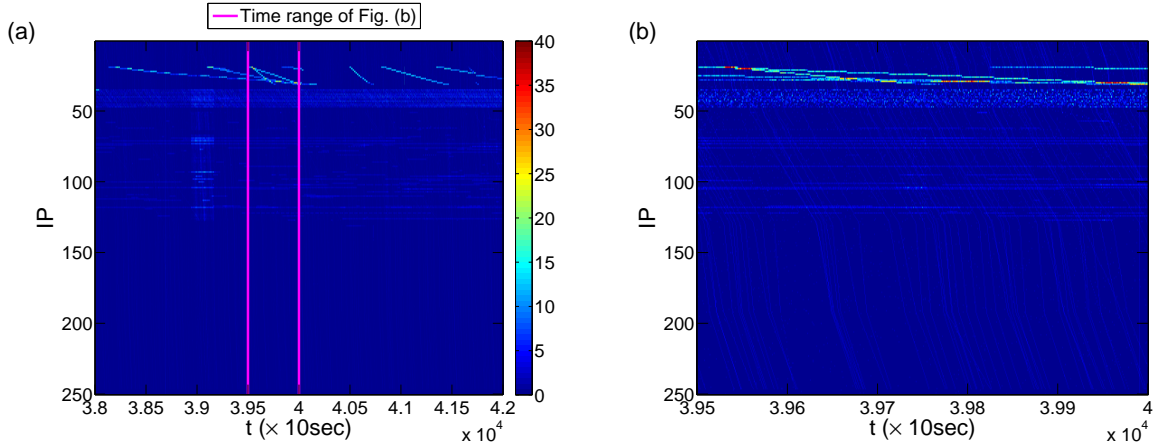

Figure 3: **Deterministic attack pattern on different time scales.** For time resolution $\Delta t = 10$ seconds and IP region 1-250, (a) deterministic attack pattern in a relatively large time scale: the sloped light-green lines in the IP region 19-31, where the time range is $t = 38000$ to $42000$, (b) enlarged section between the two vertical pink lines in (a) in the time range from $t = 39500$ to $40000$, which provides a time-scale comparison between two types of deterministic attack patterns.

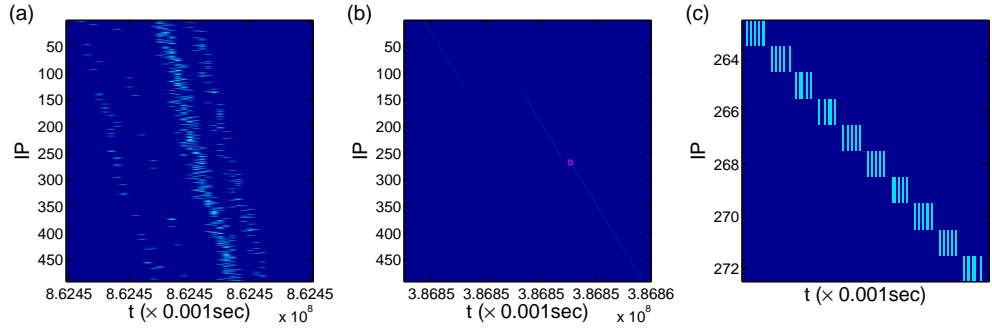

Figure 4: **Deterministic attack pattern of "walls".** For time resolution $\Delta t = 10^{-3}$ second, (a) one type of "wall" attacks over the entire IP space, with each IP receiving one attack. The time delays for the attacks to reach the targeted IP addresses exhibit three distinct values, separating the dots in this panel into three lines. (b) A different type of "wall" attack with different delay, and (c) enlarged section in the small red square in (a). Five consecutive attacks on each IP occurred before the next IP is attacked.

Our key idea is that the stochastic component of cyberattacks can be characterized as a flux distribution processes among all the IP addresses. Using the flux-fluctuation relation, we can identify and distinguish the patterns of the external drives. Fig. 5(a) shows the flux-fluctuation relation on a double logarithmic scale, where the attack frequency $w(t)$ of a victim IP corresponds to its flux, and the total number of attacks on a certain IP region is regarded as the external drive $M(t)$. We observe that a substantial part of the flux-fluctuation relation follows the scaling $\sigma \sim \sqrt{\langle w \rangle}$ (with slope $1/2$ on the logarithmic scale), while a small portion follows the scaling $\sigma \sim \langle w \rangle$ (with slope 1). These results suggest that cyberattacks share some intrinsic common features with stochastic transportation dynamics. The flux-fluctuation theory can then be used to analyze the stochastic components of cyberattack patterns.

From Fig. 5(a), we see that the flux-fluctuation relation for the most heavily bombarded IP region (35-47, see Fig. 1) displays the unity slope, indicating a non-Poisson type of external drive with strong fluctuations. The IP region under mostly deterministic attacks (IP 192-246, the aqua background with no other overlayed attack patterns) corresponds to only one dot in the flux-fluctuation diagram, establishing the deterministic nature of the attack without any randomness. Figure 5(b) shows the $\sigma$-$\langle w \rangle$ relation for the dark-blue background with no overlayed attack patterns, where we obtain the relation $\sigma = \sqrt{\langle w \rangle}$ by removing the deterministic attacks. This is mainly due to the sparsity of attacks associated with the dark-blue background. That is, without walls, mostly only one attack was received in the corresponding IP region within each time unit. We thus have $W = 1$ or 0. The average flux per unit time for one given IP is $\langle w \rangle \leq 1/N$, with $N$ being the size of the IP region. For observation with $T$ time units, the number

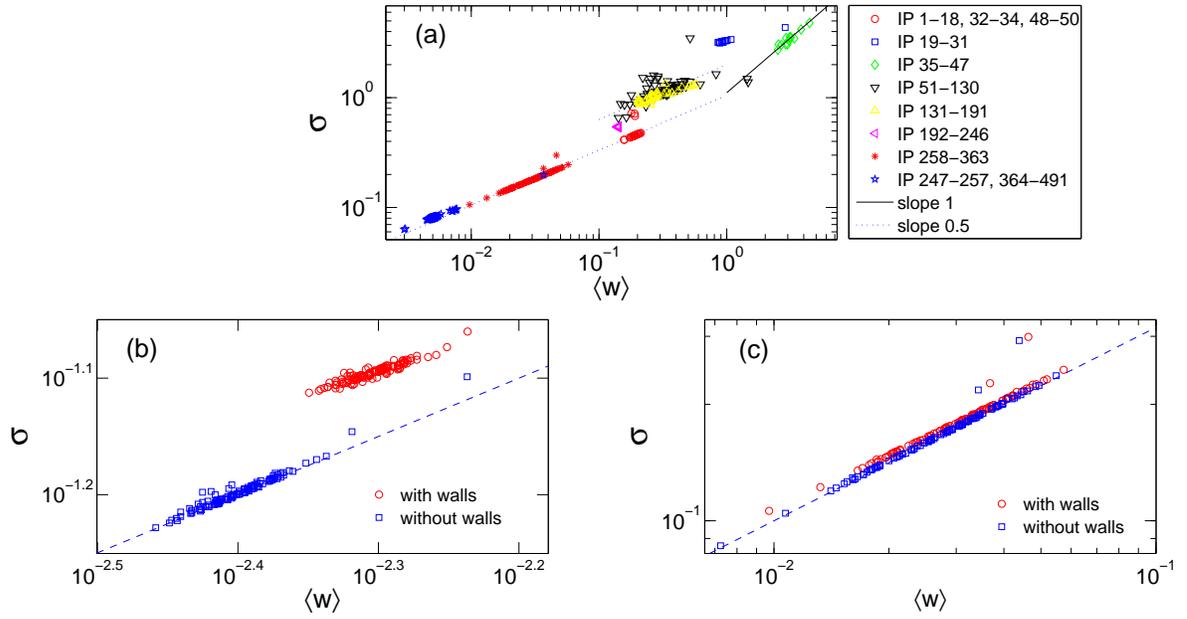

Figure 5: **Flux-fluctuation relation.** For time resolution $\Delta t = 10$ seconds, $\langle w \rangle$-$\sigma$ relations (a) for each IP in the IP space, (b) for IP regions 247-257 and 364-491, and for each IP in the dark-blue background with walls (red circles) and without walls (blue squares), (c) for IP region 258-363 with (red circles) and without (blue squares) walls. The blue dashed lines in (b) and (c) have the slope $1/2$.

of time units with $W = 1$ is denoted by $t$. We have

$$\sigma_{ext}^2 = \langle W^2 \rangle - \langle W \rangle^2 = \frac{t}{T} - \left(\frac{t}{T}\right)^2 = \langle W \rangle - \langle W \rangle^2. \tag{2}$$

Substituting this relation into Eq. (1), we obtain

$$\sigma = \sqrt{\langle w \rangle + \left(\frac{\langle W \rangle - \langle W \rangle^2}{\langle W \rangle^2} - \frac{1}{\langle W \rangle}\right)\langle w \rangle^2} = \sqrt{\langle w \rangle - \langle w \rangle^2} \approx \sqrt{\langle w \rangle}. \tag{3}$$

When deterministic attacks such as those represented by the walls are included, the fluctuation becomes

$$\sigma' = \sqrt{\frac{(t + t_{\text{wall}})}{T} - \left(\frac{t + t_{\text{wall}}}{T}\right)^2} = \sqrt{\frac{t + t_{\text{wall}}}{T} - \left(\frac{t + t_{\text{wall}}}{T}\right)^2} = \sqrt{\langle w' \rangle - \langle w' \rangle^2}, \tag{4}$$

where $\langle w' \rangle = (t + t_{\text{wall}})/T$, and $t_{\text{wall}}$ is the number of walls with one attack to each IP. Since $t_{\text{wall}} \gg t_a$, $\langle w' \rangle^2$ cannot be ignored, and so the slope of the $\sigma'$-$\langle w' \rangle$ relation on a double logarithmic scale is given by

$$k_s = \frac{\log_{10} \sigma'}{\log_{10} \langle w' \rangle} = \frac{1}{2} \frac{\ln\left(\langle w' \rangle - \langle w' \rangle^2\right)}{\log_{10} \langle w' \rangle} \leq \frac{1}{2} \frac{\log_{10} \langle w' \rangle}{\log_{10} \langle w' \rangle} = \frac{1}{2}. \tag{5}$$

As shown in Fig. 5(b), walls tend to reduce the slope slightly. This example illustrates the effect of deterministic attacks on the fluctuation of the attack flux. Similar phenomenon is observed in another IP region (IP 258-363) where a less sparse but larger $\langle w \rangle$ scale attack pattern takes place, as shown in Fig. 5(c).

_Spatial concentration of attacks._ Our analysis so far has focused on the *temporal* flux and fluctuation behavior of each IP. It is useful to study the distribution of the attacks in the IP space to distinguish the attack patterns in different IP blocks. The relation between the average attack frequency over the IP addresses in a given region, denoted by $\overline{w}_{\text{IP}}$, and the standard deviation of the attacks distributed among these IP addresses, denoted by $\sigma_{\text{IP}}$, is shown in Fig. 6, where the left and right panels correspond to the attacks associated with the dark-blue background and with IP address block 258-363, respectively. In the dark-blue background, attacks are sparse. The extreme case with $n$ attacks homogeneously distributed among $N$ IP addresses leads to $\overline{w}_{\text{IP}} = n/N$, where each of $n$ IP addresses receives $w = 1$ attack and the remaining IP addresses have $w = 0$. The standard deviation of the attacks among these IP addresses can

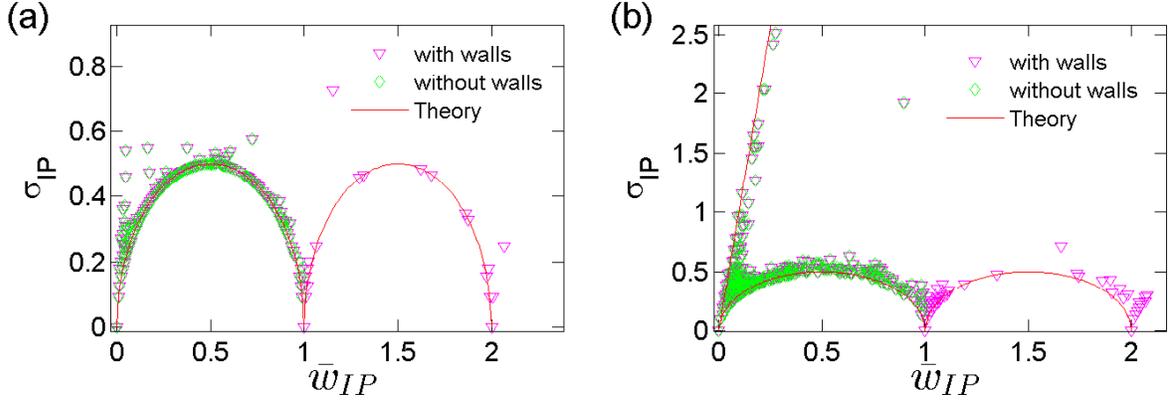

Figure 6: **Attack frequency deviation within an IP region versus the average at each time unit** for (a) the the dark-blue background and (b) IP region 258-363. The time resolution is $\Delta t = 10$ seconds.

simply be written as

$$\sigma_{\mathrm{IP}} = \sqrt{\frac{n}{N} - (\overline{w}_{\mathrm{IP}})^2} = \sqrt{\frac{1}{4} - \left(\overline{w}_{\mathrm{IP}} - \frac{1}{2}\right)^2}, \tag{6}$$

which is the equation of an ellipse. This equation matches the lower bound of the real data very well, while any degree of inhomogeneity in the attack distribution will lead to a larger value of $\sigma_{\mathrm{IP}}$. In addition, for the time resolution of 10 seconds, the effect of "wall" attacks with unity height is to increase each $\overline{w}_{\mathrm{IP}}$ exactly by 1, since a wall introduces one attack to each single IP. However, the change in $\sigma_{\mathrm{IP}}$ due to the "wall" attacks is small. Thus we observe two ellipses from real data in both Figs. 6(a) and 6(b).

The major difference between Figs. 6(a) and 6(b) is the straight line that appears only in 6(b). This is due to the concentration of multiple attacks on a few IP addresses, leaving the remaining IP addresses free of attacks. The extreme case is where all $n$ attacks focus merely on one IP. Then, the average value of $\overline{w}_{\mathrm{IP}}$ is still $n/N$, but the standard deviation becomes

$$\sigma_{\mathrm{IP}} = \sqrt{\frac{n^2}{N} - (\overline{w}_{\mathrm{IP}})^2} = \sqrt{N-1} \cdot \overline{w}_{\mathrm{IP}}, \tag{7}$$

which gives the upper bound fit (the straight line) with the real data, as shown in Fig. 6(b). Consequently, it is the concentration of attacks that distinguishes the attack pattern in the IP region (258-363) from the dark-blue background. We thus see that, for observational time resolution $\Delta t < 10$ seconds in which the attacks are sparse, the plots of $\sigma_{\mathrm{IP}}$-$\overline{w}_{\mathrm{IP}}$ relation are confined in the region bounded by Eqs. (6) and (7), and larger values of $\sigma_{\text{IP}}$ imply that the attacks are inhomogeneous in the IP space with uneven spatial concentration.

We conclude that the target selection scheme in the IP region 259-491 is highly stochastic due to the lack of regular spatiotemporal pattern. Within each time unit, there are mainly two target selection schemes: (1) all attacks are concentrated only on a single IP; (2) attacks are evenly spread over the whole region, i.e., each attack packet is received by a different IP.

*Inference probability of attack patterns.* Due to the surprisingly stable similarity of the attack patterns within an IP block, attack time sequences of frequencies on all IP addresses may be inferred from any single sequence. High inference probability of a consecutive IP region indicates that information obtained from one sensor (one IP) may be sufficient to capture the key features of the attack patterns in the whole region. Thus, our discovery of the attack pattern similarity may help to reduce dramatically the number of sensors needed to monitor the attack behavior in the whole cyberspace of interest.

A straightforward measure of inference probability is the correlation coefficient, as the time series belonging to a similar attack pattern are likely to be highly correlated. The correlation coefficient between two time series $i$ and $j$ is given by

$$\rho_{i,j} = \frac{\langle (w_i - \langle w_i \rangle)(w_j - \langle w_j \rangle) \rangle}{\sigma_i \sigma_j},$$

(8)

where $w_i$ and $w_j$ are the attack frequencies of IP addresses $i$ and $j$, and $\sigma_i$ and $\sigma_j$ are the corresponding standard deviations. Fig. 7(a) shows the correlation coefficient matrix containing $\rho_{i,j}$ for each $i$-$j$ pair with relatively low time resolution ($\Delta t = 10^4$ seconds). We see that relatively large correlation coefficients appear within the IP regions corresponding to the readily distinguishable colored blocks in Fig. 1. To further exploit the use of the correlation coefficients, we consider the IP group generated from one given IP $i$ by calculating the correlation coefficients between the time series from this IP and those from all other IP addresses, $\rho_{ij}$, over a certain threshold $\rho_c$. The sizes of the groups generated from the IP addresses under a similar attack pattern would be close to each other. From the group size of each IP shown in Fig. 7(b), we see that time series associated with IP addresses in different blocks behave differently, despite the large fluctuations. The correlation coefficient, however, may not be a reliable measure to characterize similarity and inference probability of the main attack pattern. Figure 7(c) shows a case where the correlation coefficient works well, but an unsuccessful case is shown in Fig. 7(d) where the two abrupt peaks produce very low correlation coefficients among the three time series, in spite of their similarities in regions excluding the peaks. We note that the peaks correspond to extreme events with highly concentrated attacks. When the extreme events on different IP addresses are out of phase, the correlation coefficient fails to reflect the similarity of the main attack pattern and can lead to large fluctuations. Other statistical properties, such as $M$, the total number of attacks on a particular IP, and $\langle \tau \rangle$, the average time interval between consecutive attacks, can also be used to characterize the attack pattern similarity, as shown in Fig. 7(b). We see that the fluctuations of these quantities are much smaller than $\rho_{ij}$ and they are thus able to better distinguish the IP addresses under different attack patterns. Another disadvantage of the correlation coefficients is that they are sensitive to time resolution $\Delta t$. For high resolution (small $\Delta t$), the reduction in the attack frequency values may make a time series so sparse that it is dominated by random fluctuations, and this can result in a sharp decrease in the correlation coefficients between such sparse series. For example, for $\Delta t = 10$ seconds, $\rho_{i,j}$ is close to 0 for almost any $i$-$j$ pairs.

To better characterize the attack patterns and overcome the sensitivity on time resolution and frequency value fluctuations, we coarse-grain the time series according to the order of magnitude of their amplitude, i.e., we replace each $w(t)$ by a state number, $X(t) = \text{int}[\log_{10} w(t)]$, where int denotes the integer closest to $\log_{10} w(t)$ [for $w(t) = 0$, we set $w(t) = 0.1$ so that $X(t) = -1$]. Based on $X(t)$, we can construct a Markov state transition probability matrix (MSTPM) to capture the key information about the attack pattern on any particular IP. The entry in the matrix at the $m$th row and $n$th column denotes the probability that the IP is in state $m$ at $t - \Delta t$ and it transitions to state $n$ at time $t$. Since the length of the time series is relatively large, the transition probabilities are robust against random fluctuations. In addition, for IP addresses under similar attack patterns, the similarity in their state transition patterns would hold irrespective of the time resolution. As shown in Fig. 8, IP addresses within the same attack pattern region have similar MSTPMs, even for high resolution (e.g., $\Delta t = 10$ second). This suggests that MSTPMs, which can be measured via the correlation coefficients where the matrix entries are organized into a vector according to a certain order, can be used to quantify the inference probability, as shown in Fig. 9. We observe that the IP addresses attacked under similar patterns have relatively high correlation coefficients, while low correlation coefficients can distinguish the deterministic from stochastic attacks. In general, the attack frequencies of a group of IP addresses under deterministic (or stochastic) attack can be inferred from one of them under attack of the same type.

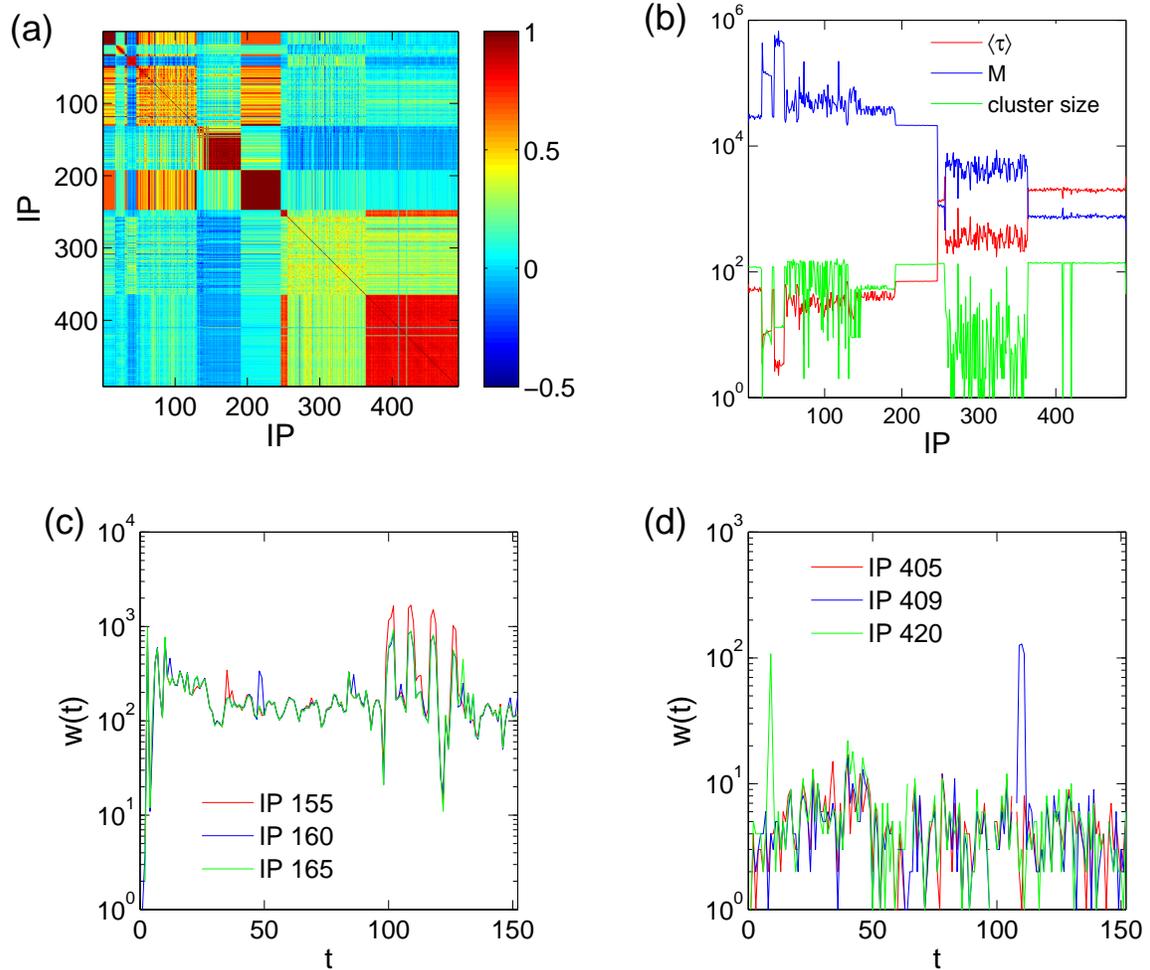

Figure 7: **Correlation coefficients**. (a) Correlation coefficient matrix associated with the IP space, (b) total number of attacks on an IP address, $M$ (blue), the average time interval between consecutive attacks on an IP, $\langle \tau \rangle$ (red), and the cluster size to which each IP belongs (green). The clustering threshold is set to be 0.7 (somewhat arbitrary). (c) Time series of attack frequency for IP 155 (red), 160 (blue), and 165 (green). All three IP addresses belong to the region of the aqua background without any other overlayed attack patterns. (d) Time series of attack frequency for IP 405 (red), 409 (blue), and 420 (green), which belong to the region of the dark-blue background without any other overlayed attack patterns. One abrupt peak occurs at IP 409, and another at IP 420. The time resolution is $\Delta t = 10000$ seconds for all four panels.

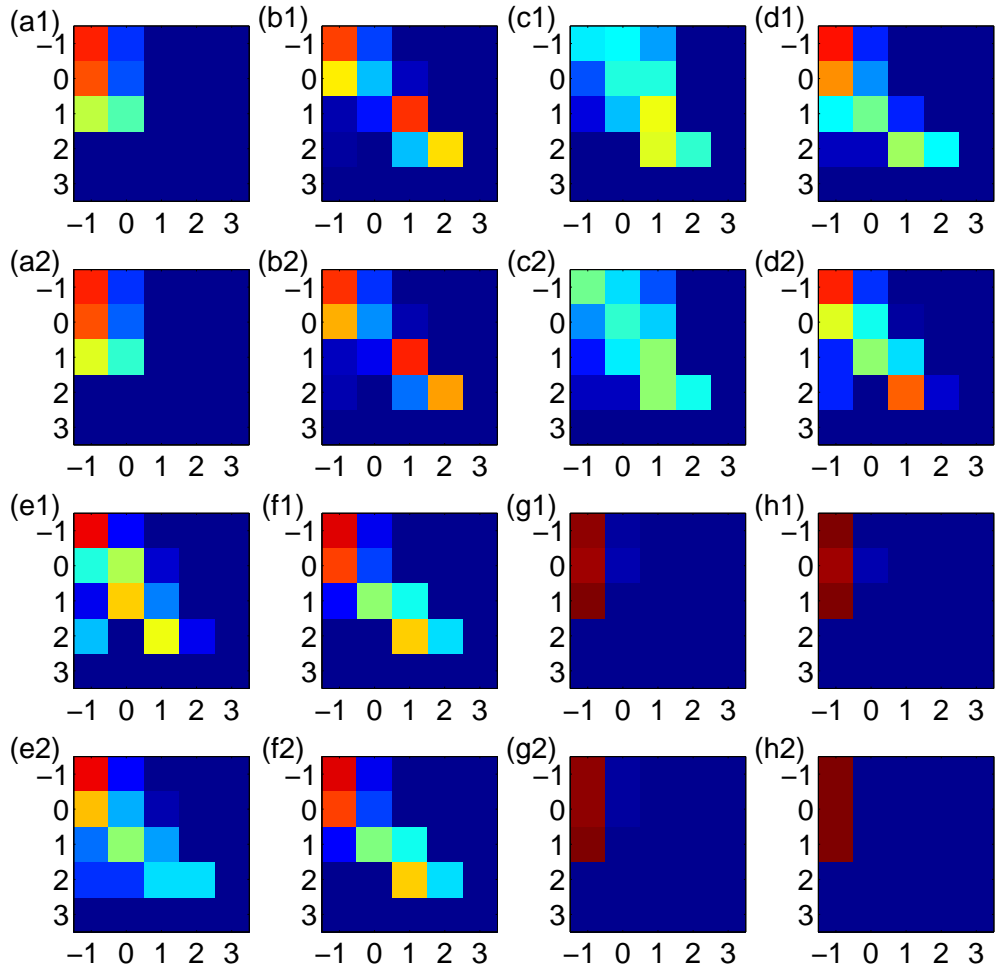

Figure 8: **Markov state transition probability matrix (MSTPMs).** For time resolution $\Delta t = 10$ seconds, MSTPMs obtained from two randomly selected IP addresses within the corresponding IP region. See the legend of Fig. 6(a) for the eight IP regions under different attack patterns: IP addresses (a1,a2) 10 and 33, (b1,b2) 20 and 30, (c1,c2) 40 and 45, (d1,d2) 60 and 120, (e1,e2) 140 and 190, (f1,f2) 200 and 240, (g1,g2) 260 and 360, and (h1,h2) 420 and 490.

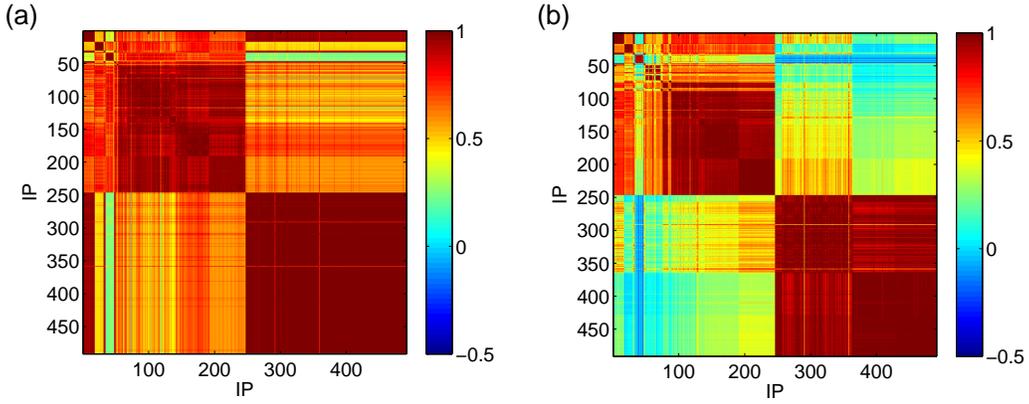

Figure 9: **Correlation coefficient matrix associated with the MSTPMs** for (a) $\Delta t = 10$ seconds and (b) $\Delta t = 100$ seconds.

Generally, attack frequency time series have high inference probability, which can be measured by the correlation coefficients or the MSTPMs. The latter, however, are more reliable especially for high time resolutions. This finding would enable one to monitor the cyberattack patterns throughout the entire IP space of interest using fewer sensors.

*Predictability of cyberattacks.* Predicting future cyberattacks is an ultimate goal in investigating cyberattack patterns through data analysis. The degree of predictability can be characterized by the uncertainty associated with the state transitions in the coarse-grained time series of attack frequencies, which can be further quantified by the information entropy. Taking into account the temporal correlations in the state transition process, we define the information entropy of IP $i$ as

$$E_i = - \sum_{S_i' \subset S_i} P(S_i') \log_2[P(S_i')], \tag{9}$$

where $S_i = \{X_1, X_2, \ldots, X_T\}$ denotes the sequence of states that the time series reached for IP $i$, and $P(S_i')$ is the probability that the state sequence $S_i'$ appears to IP $i$. The information entropy $E_i$ takes into account both the heterogeneous probability distribution in different states and the temporal correlations among the states. The entropy is thus able to provide a realistic characterization of the attack patterns.

The predictability $\Pi$ of a state sequence is defined as the success rate that an algorithm can achieve in predicting the sequence's future states [58]. For a sequence with $N_S$ possible states, the predictability measure is subject to the Fano's inequality: $\Pi \leq \Pi^{\max}(E, N_S)$, where the predictability upper bound

$\Pi^{\mathrm{max}}(E, N_{\mathrm{S}})$ is obtained by solving the following equation

$$E = -\Pi^{\mathrm{max}} \log_2(\Pi^{\mathrm{max}}) - (1 - \Pi^{\mathrm{max}}) \log_2(1 - \Pi^{\mathrm{max}}) + (1 - \Pi^{\mathrm{max}}) \log_2(N_{\mathrm{S}} - 1). \qquad (10)$$

In the predictability calculation, high time resolutions, e.g., $\Delta t = 10$ seconds, can make the series so sparse that the overwhelming majority of the states are reduced to the "ground state" $X(t) = -1$ [corresponding to $w(t) = 0$]. In this case, the predictability $\Pi^{\mathrm{max}}$ assumes artificially high values, as the future states can trivially be predicted to be $X(t) = -1$, leading to a high success rate. To avoid this artifact, we need to use moderate time resolutions, e.g., $\Delta t = 100$ seconds or 1000 seconds. The length of the time series can affect the value of the predictability. Predictable patterns may not be fully contained within short time series and, hence, we evenly divide the whole time series into multiple sections, each of length of $h$ hours. Figures 10(a) and 10(b) show the predictability upper bound $\Pi^{\mathrm{max}}$ for different section lengths and time resolutions. While longer time series (higher $h$) tend to be more predictable as expected, all time series have surprisingly high predictability. There are cases where the predictability probability is over 90%, indicating that it is possible to make correct predictions for over 90% of the cases. We observe that the IP regions of different attack patterns in Fig. 1 have different $\Pi^{\mathrm{max}}$ values, and the IP addresses under each attack pattern are approximately equally predictable. Deterministic attacks have higher predictability than stochastic attacks for $\Delta t = 1000$ seconds, but the opposite occurs for $\Delta t = 100$ seconds, due to the fact that stochastic attacks are much more sparse and, as such, a higher time resolution can introduce overwhelmingly more ground states into the state sequence. Despite the different attack patterns, the average predictability over all IP addresses converges to about 93% with relatively small error bars as the section length is increased, as shown in Fig. 10(c). Figure 10(d) shows the average predictability of the deterministic and stochastic attacks. We see that deterministic attacks systematically have higher predictability than stochastic attacks for $\Delta = 1000$ seconds. Again, sparsity contributes to the relatively high predictability (over 85% for long sections) of stochastic attacks.

## Discussion

*How do attackers choose their targets?* From Fig. 7(b), we see that the IP addresses under a similar attack pattern have similar $M$ values, which means that different IP addresses in the IP region dominated by one particular attack pattern are approximately equally visited by the attackers. Due to this, we speculate that in most cases, an attacker does not target some specific IP addresses, but a consecutive IP region. In the case of deterministic attacks, the attacker scans a whole block of adjacent IP addresses by using its botnet, and each IP in the targeted block has equal probability to be hit. In the case of stochastic attacks, at each time an attack is launched, IP addresses in the targeted block have similar chances to receive it. It may indicate that the attackers are most likely to view the targeted IP block as a black-box, the internal structure of which may be irrelevant.

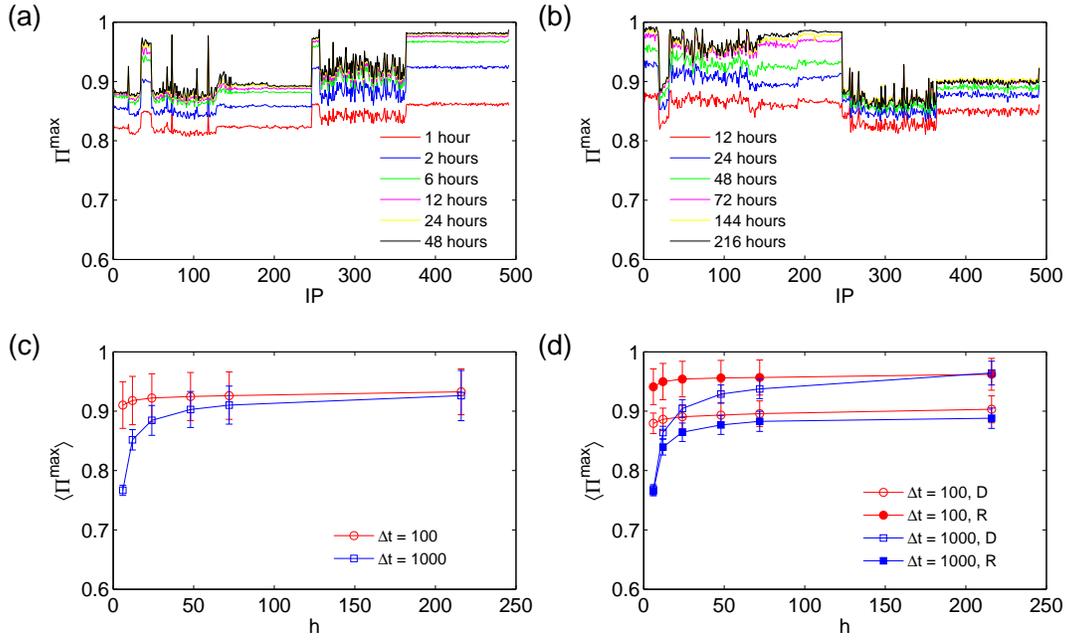

Figure 10: **Predictability calculated from coarse-grained time series.** (a,b) Predictability upper bound $\Pi^{\max}$ for all 491 IP addresses for various section lengths under time resolutions $\Delta t = 100$ seconds and 1000 seconds, respectively, where $\Pi^{\max}$ values are averaged over all sections. (c) Average $\Pi^{\max}$ versus section length $h$ for $\Delta t = 100$ seconds (red circles) and 1000 seconds (blue squares). (d) Average $\Pi^{\max}$ of the deterministic (open symbols) and stochastic (closed) attacks versus the section length $h$ for $\Delta t = 100$ seconds (red circles) and 1000 seconds (blue squares). In the figure, "D" denotes deterministic and "R" stands for random or stochastic. See text for explanation on why, in the case of $\Delta t = 100$, stochastic attacks can have a higher predictability upper bound than deterministic attacks.

*Interplay between different attack patterns.* Except for the parts in Fig. 5 that show slope 1 or 1/2, there are some parts of the data that do not exhibit a plausible scaling, which correspond to the high attack density patterns that are overlayed on the deterministic aqua background. Figs. 2(a) and 3(b) show that the light-blue lines change their slopes when other attack pattern are superimposed on them. This reveals an interesting phenomenon: different attack patterns may interfere with each other. If so, it would be of great importance to investigate how a certain type of attack pattern promotes or suppresses the efficiency of another type, and how the attack packets differ from normal data packets when transporting through the Internet. Making use of the suppression effect can lead to a dramatic reduction in the transportation efficiency of the attack packets. Our work indicates that it is potentially possible to mitigate or eliminate cybersecurity threats substantially through a macroscopic approach, a field that requires much further efforts.

*Attack pattern inference and prediction.* The surprising finding that cyberattack patterns are highly predictable encourages us to develop inference and prediction algorithms. The former can provide us with global insights into cybersecurity based on limited information resources, and the latter would enable us not just to get a better understanding of current cyberattack data but also help us to forecast future cyberthreats.

*Possible improvement for future IDS.* The analysis on the deterministic attack reveals the possibility for an IDS to be prepared in advance to the coming attack based on the estimation of the relatively constant attack frequency. If we could make the IDSs in a consecutive IP region communicate on attack information, then the sweeping speed can be estimated, which more accurately tells each IDS when shall the next attack take place, since the adjacent victim IPs are swept by the attacker one after another in order within a constant time interval.

*About the IP information of the attackers.* This paper focuses on the analysis of data obtained via the victim side. In addition, with the IP information of the attackers, if possible, we would be able to further verify our speculation that each sweeping is generated from a single IP address. The spatial distribution of the attackers' IPs can also be retrieved. For deterministic attack, the sweepings are targeted on consecutive IPs, and it is very likely that a substantial part of the attackers' IPs are also consecutive or nearly consecutive, since those attacker could be ex-victims of the same type of attack, which have been successfully compromised by a major attacker and are now serving as a part of its botnet. More properties of cyberattack which can not be observed on the victims' side could also be revealed via such information.

## Materials and Methods

The data set we analyzed was collected between 2/9/2011 and 2/25/2011 by a cyber instrument known as honeypot (see http://www.honeynet.org). The data set contains 491 honeypot IP addresses. The honeypot simulates vulnerable computer services corresponding to distinct TCP ports. Since there are no legitimate services associated with these IP addresses, the traffic arriving at these ports is widely deemed as attacks. The raw network traffic arriving at these ports was initially recorded as `pcap` files. By using some standard pre-processing procedure, we reformulated the raw traffic into *flows*, which are the commonly accepted representation of cyber attacks. The pre-processing procedure uses two widely used parameters: the *flow timeout time* of 60 seconds, meaning that a flow expires after 60 seconds of no more packets arriving activities; the *flow lifetime* of 300 seconds, meaning that a flow expires after 300 seconds. The supplementary material contains the data set, which has two columns. Each row is a tuple $(a, b)$, which represents that IP address $a$ was attacked at time $b$ at some TCP port. Since our analysis treats an IP address as a whole, we do not distinguish the specific ports. For the purpose of protecting privacy, the honeypot IP addresses (i.e., the first column) were anonymized via a one-to-one mapping that also maintains their consecutiveness.